\documentclass{vgtc}                          

\ifpdf
  \pdfoutput=1\relax                   
  \usepackage{graphicx}                
  \DeclareGraphicsExtensions{.pdf,.png,.jpg,.jpeg} 
\else
  \usepackage{graphicx}                
  \DeclareGraphicsExtensions{.eps}     
\fi%

\graphicspath{{figures/}{pictures/}{images/}{./}} 

\usepackage{microtype}                 
\PassOptionsToPackage{warn}{textcomp}  
\usepackage{textcomp}                  
\usepackage{mathptmx}                  
\usepackage{times}                     
\usepackage{cite}                      
\usepackage{tabu}                      
\usepackage{booktabs}                  

\onlineid{0}

\vgtccategory{Research}

\vgtcinsertpkg

\title{\emph{Music-Circles}: Can Music Be Represented With Numbers?}

\teaser{
 \includegraphics[width=\textwidth]{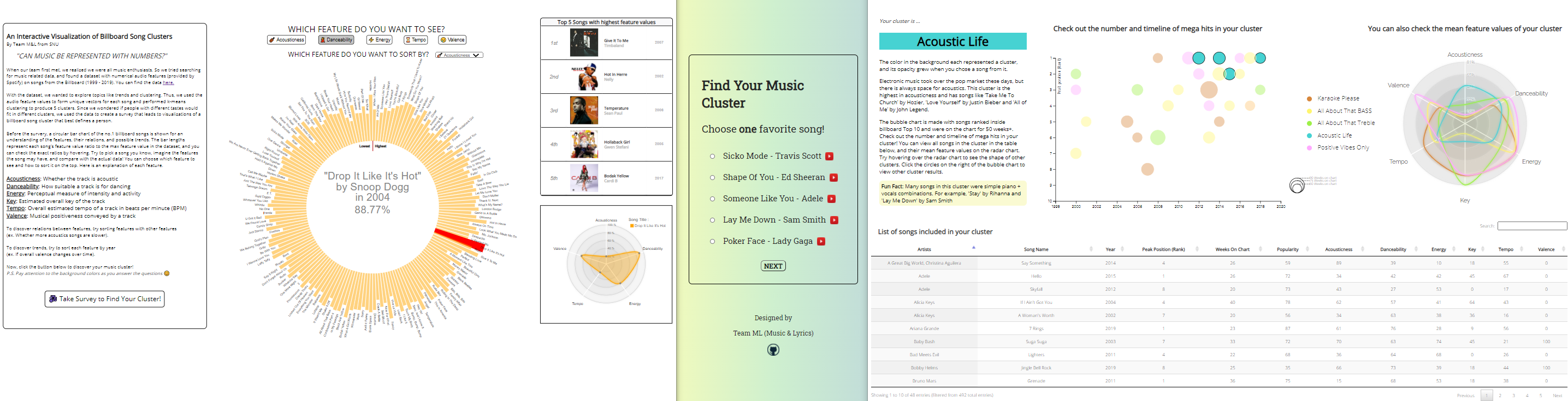}
 \caption{Overview of the three pages of \emph{Music-Circles}. Left: First Page (Circular Barplot, Table, Radar Chart); Middle: Second Page (Survey); Right: Third Page (Cluster Visualization - Bubble Chart, Radar Chart, Table)}
 \label{fig:overview} 
}

\author{Seokgi Kim, Jihye Park, Kihong Seong, Namwoo Cho, Junho Min, Hwajung Hong\thanks{Corresponding author.}
\thanks{Date of current manuscript version January 4, 2021. {\emph{(All authors contributed equally to this work.)}}
\newline
S. Kim, K. Seong, J. Min are with the Ambient NLP Laboratory, Graduate School of Data Science, Seoul National University, South Korea (e-mail: blaqdraq77@snu.ac.kr; kev94yo@snu.ac.kr; snu11mjh@snu.ac.kr).
\newline
J. Park is with the Data Mining Laboratory, Department of Industrial Engineering, Seoul National University, South Korea (e-mail: jihyeparkk@snu.ac.kr). 
\newline
N. Cho is with with the Human Interface Systems Laboratory, Department of Industrial Engineering, Seoul National University, South Korea (e-mail: chonamwoo@snu.ac.kr). 
\newline
Code is available at: \href{https://github.com/chonamwoo/snuDxd}{\texttt{https://github.com/chonamwoo/snuDxd}}
}}

\abstract{
The world today is experiencing an abundance of music like no other time, and attempts to group music into clusters have become increasingly prevalent. Common standards for grouping music were songs, artists, and genres, with artists or songs exploring similar genres of music seen as related. These clustering attempts serve critical purposes for various stakeholders involved in the music industry. For end users of music services, they may want to group their music so that they can easily navigate inside their music library; for music streaming platforms like Spotify, companies may want to establish a solid dataset of related songs in order to successfully provide personalized music recommendations and coherent playlists to their users. Due to increased competition in the streaming market, platforms are trying their best to find novel ways of learning similarities between audio to gain competitive advantage. Our team, comprised of music lovers with different tastes, was interested in the same issue, and created \emph{Music-Circles}, an interactive visualization of music from the Billboard. \emph{Music-Circles} links audio feature data offered by Spotify to popular songs to create unique vectors for each song, and calculate similarities between these vectors to cluster them. Through interacting with \emph{Music-Circles}, users can gain understandings of audio features, view characteristic trends in popular music, and find out which music cluster they belong to. 
} 

\CCScatlist{
  \CCScatTwelve{Music Clustering}{K-Means Clustering}{Mu\-sic Re\-com\-mendation}{};
  \CCScatTwelve{Data Visualization}{Interactive Visualization}{}{}
}

\begin{document}


\firstsection{Introduction}

\maketitle

The music industry went through significant adjustments whenever the prominent medium used to listen to music changed. Three major changes in medium for music are notable; Vinyls to CDs, CDs to digital downloads, digital downloads to online streaming. The second change from CDs to digital downloads introduced a monumental shift of music from tangible forms to digital forms, and the third transition to online streaming led to a demand for enhanced analytics of the digitized audio content. This was because online streaming platforms offered the complete music library to listen to by a subscription basis, whereas in the era of downloads, users would find songs that they like and download them to listen. Given a sea of audio contents, users might get lost, so the streaming platforms decided to use personalized music recommendations as a guide provided to the users.

To make this guide, platforms had to invent an algorithm that finds similar audio contents to a song that a certain user likes. Therefore, through various methods such as collaborative filtering and content based filtering, users are recommended songs that they might like. However, in general, the algorithm itself isn't explained to the users, and the recommendations take forms of song, artist and album names with album covers provided if available. If the user has no context about the recommended song, artist and album, there is no visual aid to help the user understand that this song is actually related to other songs that the user prefers. 

In this paper, we present \emph{Music-Circles}, a system that supports user comprehension of audio features used to cluster music, and offers visualizations of different music clusters. \emph{Music-Circles} bring together two music related data sources with several audio features and billboard chart song information into integrated visual interfaces to help users understand, and offer coordinated visualizations that change simultaneously as the user interacts. The coordinated views are expected to give the users macro and micro pictures of different features and clusters. This multiview style look was selected since we wanted to create a functional system that can help clearly understand the data. Fig.~\ref{fig:overview} shows an overview of the system. The first page, which we will refer to as the discover-screen in the sections below, contains annotations and multiple visualizations that will give the user an understanding of the audio features of songs. Then the system moves on to the second page (will be referred to as the survey page), which contains a survey that will help the users find a music cluster that matches their taste. After completing the survey, users are directed to the third page (will be referred to as the cluster-view), which shows annotations and multiple visualizations about the personalized cluster results. \emph{Music-Circles} uses songs from the Billboard hot 100 songs chart so that there will be a high chance of songs being familiar to users, and allows users to analyze these songs in ways that serve the users interest. For example, the user can see what audio features are similar between two songs or find out a certain feature value for a familiar song to get an understanding of which musical feelings/vibes lead to which features. 
\\
Main contributions of our work are as follows:
\begin{itemize}
    \item A survey research in the field of music consumption;
    \item \emph{Music-Circles}, an interactive visual analytics system for exploring audio feature trends of popular songs and discovering characteristics of personalized song clusters;
\end{itemize}

\section{Related Work}
Primary purpose of music is entertainment and activities surrounding music including music discovery should be engaging and entertaining; the main purpose of our work is to create an interesting visualization system that music enthusiasts can enjoy \cite{lamere2007explore}. Thus, we have referred to several works to effectively visualize music data.

To visualize individual characteristics of each song, a few issues should be taken into account. First, each song should be displayed in an equal relationship rather than a vertical relationship. Second, it should be possible to put numerous songs on one screen. Although Guedes and Freitas \cite{guedes2017exploring} solved these issues by arranging songs in a circular shape, there still seemed to be room for improvement since they did not include sorting buttons that can sort the features.

The main data of our work are features of the songs. Film Flowers\footnote{\url{https://sxywu.com/filmflowers/}} visualizes their movie data in terms of the size of flowers, number of petals, and colors. Inspired by this, we came up with a visualization to simultaneously show the data for each song. By using the radar chart, we succeeded in simultaneously expressing features of data for songs and data for each cluster.

You can see the multiview style in both the discover-screen and the cluster-view. We referred to PlayerView\footnote{\url{https://buckets.peterbeshai.com/}} and confirmed that more information can be easily obtained by connecting each chart inside a multiview system. The discover-screen includes a circular barplot, radar chart, and a top songs table to form a multiview of the feature information of each song. In the cluster-view, a bubble chart, radar chart, and an interactive table containing songs about each cluster form a multiview of the information about the cluster.

\section{Task Abstraction}

Eliciting tasks and system requirements are not easy in visualizing music since users usually cannot articulate their needs directly. Therefore, rather than simply introspecting about the needs and actions taken in exploring information within the music, interviews were utilized, since an interview is the arsenal of ethnographically inspired methods used in human computer interaction \cite{munzner}.

After interviewing five people who listen to music on a daily basis with great interests in music, we identified some of their difficulties and needs. 
Music enthusiasts were often interested in the information about the songs that they like. Yet, the current practice is that they arbitrarily receive one-sided information of relevant songs chosen by the algorithms used in the streaming services unless they proactively search for the information themselves. They wanted an easier way to find more about the patterns and features within the songs that they like and within the popular songs.

We then set two goals of our visualizations; enabling the users to readily see the trends by different characteristics of the songs and song clusters, and allowing them to freely explore the given visualizations to find out patterns of their interest with interactions and annotations provided as a guide for these interactions.
We also determined the actions the user could take: sorting the features and thus analyzing trends across all features, spotting outlier values, finding out songs matching one's taste, etc. Furthermore, the team decided to enhance engagement and enjoyment for users during these actions. These analyses became the rationales behind determining appropriate things to visualize, settling the strategies for handling the complexity within the data, and choosing effective visual idioms.

\section{Data Abstraction}

\subsection{Data Collection}
We collected two datasets from Kaggle. One\footnote{\url{https://www.kaggle.com/danield2255/data-on-songs-from-billboard-19992019}} contains data on artists and songs that appeared on Billboard Hot 100 from 1999 to 2019; and the other\footnote{\url{https://www.kaggle.com/yamaerenay/spotify-dataset-19212020-160k-tracks}} offered by Spotify contains audio features of 160,000 or more songs released in between 1921 and 2020. These audio features were acquired by Spotify through surveys with 10,000 or more listeners on their affective experience of the songs; such lexical Kansei parameters are very useful for sentimental analysis and affective attribute classfication \cite{Want2019kansei}.  

\subsection{Data Transformation}

There are hot 100 ranking columns in the Billboard dataset, and numerous features in the Spotify dataset. We merged these two tables, linking audio features in the Spotify dataset to Billboard songs by song name, to form one combined dataset containing songs from the Billboard and their audio feature values.

To identify groups of similar songs, we chose the K-means clustering method. Prior to applying clustering method, we performed several preprocessing steps on the audio features. First, we identified duplicate songs based on the title and artist and removed songs except for songs with the highest weekly rank among the duplicate songs. The total number of songs used in this study was 4,314. Second, we excluded four binary-features among the 14 audio features, since they were deemed unfit for clustering. Third, we normalized the range of three features (key, loudness, tempo) ranging from 0 to 1 in order to match the range of other features.

After exploring different clustering options in terms of features and number of clusters, we carefully chose six features (acousticness, danceability, energy, key, tempo, valence) for five clusters. The final clusters seemed to be easily distinguished from one another without having extremely low or high feature values. 

\section{Visual Design}
After the data was transformed to suit our needs, it was visualized as several idioms to serve the purpose of \emph{Music-Circles}.

\begin{figure}[t]
\center
\includegraphics[width=\columnwidth]{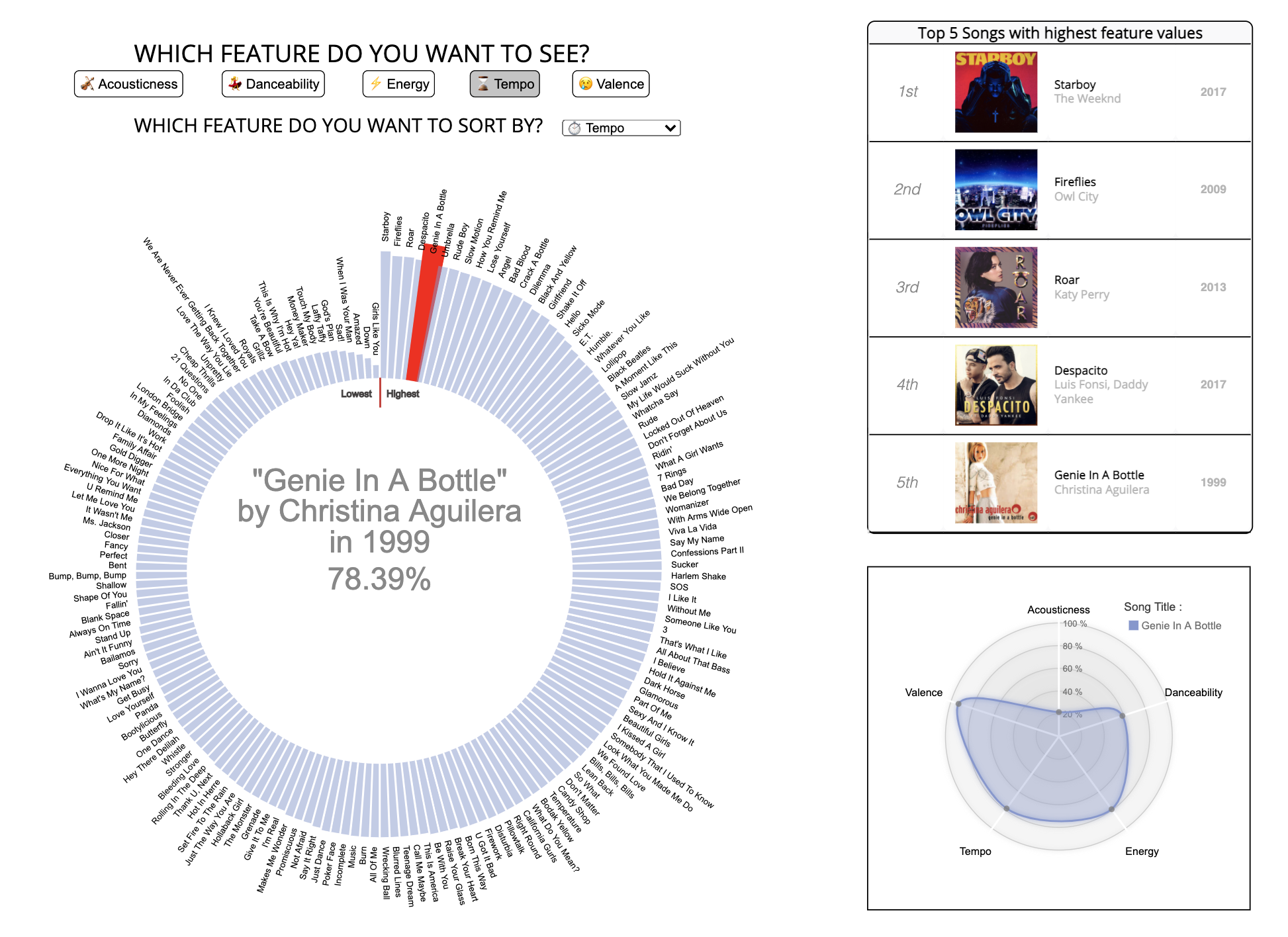}
\caption{Discover-Screen: Circular Barplot, Top 5 Table, Radar Chart}
\label{fig:circular-barplot}
\end{figure}
\subsection{Discover-Screen}
The discover-screen is the initial page presented to the user, and is intended to help users gain insight about what the audio features mean, and possibly discover trends and relations between the features. Annotations shown in the left image of Fig.~\ref{fig:overview} are included as a guide to help users understand our intentions, and easily interact with the visualizations.

\subsubsection{Circular Barplot}
Circular barplots represent data by arranging bars in a circle. A snapshot of the discover-screen, Fig.~\ref{fig:circular-barplot}, shows such visualization implemented in our work. With the intention of examining the characteristics of popular songs, we visualized 168 songs that ranked No.1 on the Billboard chart. The length of each bar means the value of the selected feature of the song, and the title of each song is marked at the end of the bar.

\subsubsection{The Feature Buttons}
Users can select 5 kinds of features (Acousticness, Danceability, Energy, Tempo, and Valence) under the 'Which feature do you want to see?' annotation in the top. When the feature button is pressed, the new circular barplot appears as a feature corresponding to the button. We designed the color of the bars to change when a new feature is selected. This animation was included in order to make the interaction process more interesting. Users can also see the 'Top 5 Songs Table' displayed at the top right, which shows the 5 songs of the highest value for the selected feature including the album images. With this table, users are expected to relate characteristics of songs they know to certain features to get a better understanding of what the features mean. For example, the table in Fig.~\ref{fig:circular-barplot} shows that the song 'Despacito' was among the songs with the highest tempo value. Users who are familiar with this song can get a better feel of what a high tempo song would sound like.

\subsubsection{The Sorting Button}
With the intention of offering users the capability to examine the trend of relationships between one feature and others, we implemented a sorting button in the form of dropdown options, which can sort the bars according to the 5 different features. When a sorting option is selected, sorting is performed so that the bars are sorted in clockwise direction, from songs with the highest selected feature values to songs with the lowest values. The reason why we chose to sort from the highest to the lowest in clockwise direction is because we assumed that users would probably want to see the song with the highest value first when they select a feature. Since we marked the highest and lowest point, users can easily recognize which song has the highest or lowest value.

\subsubsection{Hoverable Bars}
All bars are hoverable, and enables users to see information of the song (title, artist, release year, and the selected feature value) in the middle when the cursor its hovered on them. A radar chart which simultaneously reacts to this hovering action in the bottom right of Fig.~\ref{fig:circular-barplot} shows the 5 feature values of the selected song, so that the overall feature distribution of each song can be seen.

\subsubsection{Radar Chart}
\label{sec:radar}
Radar charts are one of the useful ways to display multivariate observations with an arbitrary number of variables; thus, we use a radar chart to display various audio features of each song. To prevent users from being overwhelmed by the visual clutter, we display less than 7 features, which abides by the Miller’s magic number seven \cite{miller1956magic}. 

With dots spotted on each radial position along the axes representing each attribute of the component, users can easily grasp the selected song’s characteristics. 
As mentioned in the explanation of hoverable bars, the values in the radar chart change as users hover over the bars and explore different songs. Furthermore, the color of the chart changes as users click on the feature buttons to examine different features; the colors were carefully selected to not hinder the apprehension of the features, and be easily distinguishable. 

\begin{figure}[t]
\center
\includegraphics[width=\columnwidth]{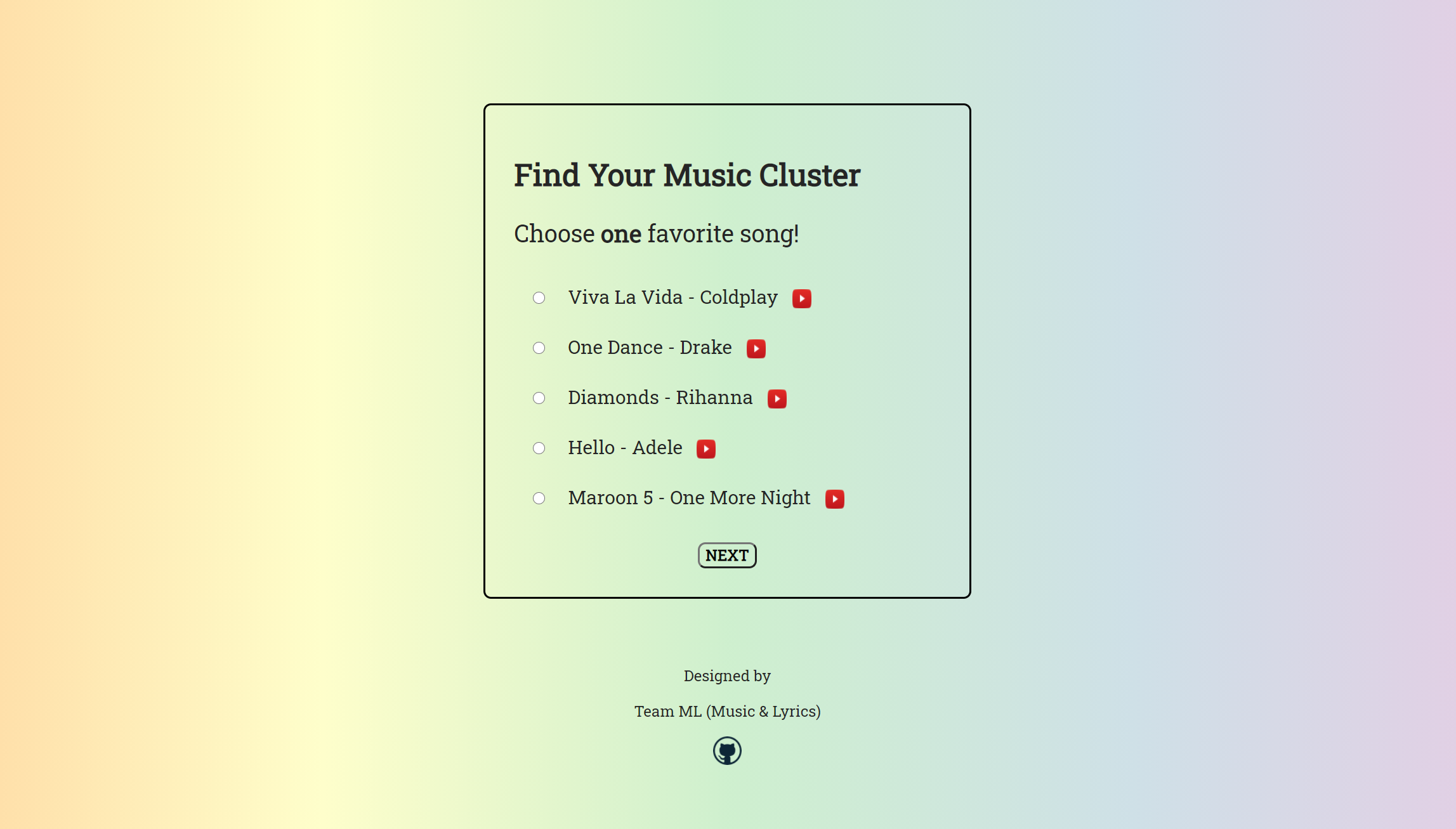}
\caption{Survey Page}
\label{fig:survey}
\end{figure}
\begin{figure}[t]
\center
\includegraphics[width=\columnwidth]{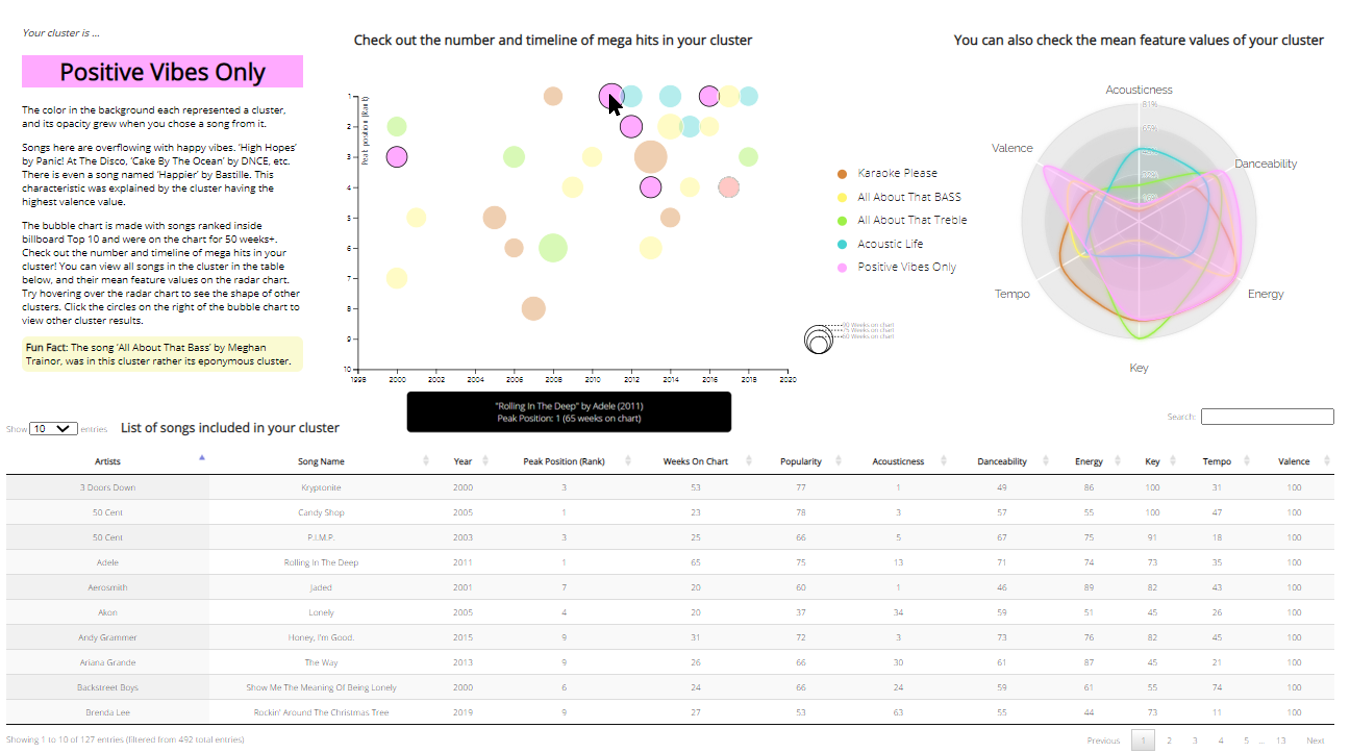}
\caption{Cluster-View: Visualizations of information about the cluster assigned to each user after survey}
\label{fig:cluster-view}
\end{figure}
\subsection{Survey}
When users are done with exploring the overall trends and characteristics of the audio features in the discover-screen, a button at the end of annotations shown in the bottom left of the left most image in Fig.\ref{fig:overview} guides them to a survey page. This survey page is intended to enhance user involvement and thus make our project more interesting. By completing the survey, users can discover what clusters their tastes belong to, and explore the clusters' characteristics. The survey page design is a simple survey outline, with one question asking to choose one favorite song, five song options, and a 'next/finish' button to move on to the next question. The five songs are each selected from one of the five clusters that we defined in Fig.\ref{fig:cluster-view}, and songs deemed as most renowned, like Coldplay's 'Viva La Vida' or Drake's 'One Dance' are presented so there would be a high chance that a user is familiar with the song. If the user does not recognize a song, a youtube link in the form of an icon is presented for the user to visit and listen to the song.

The background is a left to right color gradient with 5 colors, each corresponding to one of the five predefined clusters. For example, the orange color (far left) in Fig.\ref{fig:survey} corresponds to cluster 0 in Fig.\ref{fig:cluster-view}. The colors are selected are from a rainbow palette. Although we initially wanted to use a different palette, after some experimentation we found the rainbow palette to be the most effective in distinguishing between clusters. To make the colors easing to the eye, we chose light, pastel versions of the colors. When the user chooses a song in a certain cluster, the color corresponding to that cluster increases in opacity. This animation was meant to intrigue the users while they are taking the survey, and make the process less dull. The user is not informed about the nature of the background colors, and only told to pay attention to the background through an annotation in the discover-screen. This was for the user to wonder about the animations in the background and be more immersed in the whole process.

After choosing four songs from a series of four questions, the user is presented with a 'View Results' button that will lead to a page containing the visualizations of a certain cluster. We chose to include four questions only because we did not want the survey process to take too much time, and wanted to focus on the actual visualizations. With the chosen four song names, a Javascript code is designed to fetch the matching rows from the dataset containing song feature vectors, and compute one mean vector. Then, the mean vector will be compared to the mean feature vectors for the five clusters with cosine similarity (a common way of deriving similarity between vectors), to find out which cluster the user belongs to. Such similarity models basing on closeness of distance function is widely used in generating playlists \cite{knees2006interface}.

\subsection{Cluster View}
After completing the survey, users are presented with the cluster-view, where they can explore the cluster assigned to them through three visualization idioms: bubble chart, radar chart, and table. The three visualizations simultaneously change as users click on the legends in the center, showing the difference between clusters. These linked visualization idioms provide a coordinated view of different perspectives. An annotation is provided in the top left of the page, containing explanations about the cluster based on characteristics of feature values, and a guide on what the visualizations mean and how to interact with them. A name is assigned to each cluster to make the engagement process more entertaining and cluster characteristics easier to understand. The five names are; Karaoke Please, All About That BASS, All About That Treble, Acoustic Life, Positive Vibes Only. One fun fact per cluster is also added to make the reading much more enjoyable. The overview of the visualization for cluster information for the 'Positive Vibes Only' cluster is illustrated in Fig.~\ref{fig:cluster-view}.

\subsubsection{Bubble Chart}
Bubble charts, which represent data by circles of different sizes and colors, are one of the best ways to simultaneously visualize trends in three or more dimensions \cite{robertson2008effectiveness}. As shown in Fig.~\ref{fig:cluster-view}, we propose to visualize the trend of mega-hit songs in multiple dimensions through an interactive bubble chart. Each bubble represents a mega-hit song, which was not only ranked inside the Billboard Top 10 at least once but also included more than 50 weeks in the Billboard Hot 100. The x-axis, y-axis, size, and color represent the year the song was released, peak position at Billboard Hot 100, number of weeks listed on Billboard Hot 100, and the category of a cluster, respectively. Users can also identify the details of the song with a tooltip by hovering over each bubble. Furthermore, the legends are clickable, allowing the users to explore the identities of clusters other than the cluster assigned to the user.

\subsubsection{Radar Chart}
As discussed in Section~\ref{sec:radar}, radar charts are one of the useful graphical methods to display multivariate values of the audio features. The radar chart next to the bubble chart displays the average of the audio feature values of the songs in each cluster, allowing users to grasp the identity of each cluster.

\subsubsection{Interactive Table}
The table at the bottom of the page enables effective browsing for the songs. Users can not only sort the table in lexicographic order for each column by selecting the table column heading but also search for a specific song title or artist.

\section{Results}
We developed \emph{Music-Circles} to make exploring the features of popular songs possible for the entertainment of music enthusiasts. To achieve this goal, \emph{Music-Circles} consists of three parts.

In the first part, discover-screen, through the circular barplot in Fig.~\ref{fig:circular-barplot}, 5 kinds of features of 168 songs can be compared through the length of bars. There are feature buttons and a sorting button that make users able to see different feature information or to sort the songs by some given feature. Additionally, a top 5 songs table was created to make it easier for users to see the songs with high values of a feature at one glance.

After discovering the features of these popular songs, the second part, survey page is presented. This page contains a survey to discover a users' musical tendencies as shown in Fig.~\ref{fig:survey}. During the survey, the background color changes according to the song the user chooses so the process would not be dull. Users can also get a chance to listen to the music when they find unfamiliar songs through the youtube links attached to each song.

After the survey, the last part is the cluster-view. When the users click the 'View Results' button, the cluster-view screen, as shown in Fig.~\ref{fig:cluster-view}, appears differently according to the users. Users can resonate as they read the explanations of the selected cluster, and also see which songs are included in the cluster. Also, it can help them decide which song clusters to check out in the future by looking at the radar chart displayed differently depending on the cluster.

\section{Discussion and Future Work}
Our work presents a new interactive way to explore the identity of music through its attributes. Users may obtain additional knowledge than what they would initially obtain from just the Billboard rankings or Spotify audio features. Our work encourages users to be highly involved through interactive circular barplot, survey assigning different clusters to users, and annotations with interesting facts. An important finding in our study is that annotation plays an essential role in conveying key points in visual data-driven storytelling; it helps presenters explain and emphasize core messages and specific data \cite{ren2017chartaccent}.

Without access to the raw audio signals, the scope for the audio features used in this work is limited to those that can be derived from the Spotify API. Our work could be improved by future investigations and comparisons on audio features from other sources such as the Million Data Song dataset \cite{bertin2011million}. 

\section{Conclusion}
This paper introduces \emph{Music-Circles}, an interactive system with annotations, visualizations, and surveys to keep users entertained while they engage with the system to explore audio features and clusters of popular songs. Idioms of visualizations were original ideas from the team and influenced by related works, and a focus group interview was conducted to define our task. Data from Spotify and Billboard were transformed to well suit the purpose of our project, and the transformed dataset was used to provide a coordinated view of multiple visualizations like the circular barplot, bubble chart, and radar chart to allow users to well analyze the given data.

\emph{Music-Circles} is mainly meant for entertainment of music lovers, but can be extended to other domains for different users. Producers can use our system to discover feature trends in popular songs and gain better insight in developing a hit song, and journalists can use various analyses obtained from our system to write articles on pop culture. Most importantly, \emph{Music-Circles} will fill the lack of visualizations offered in music recommendation platforms, and help users better understand characteristics between similar songs. 


\acknowledgments{
The authors wish to thank Prof. Hwajung Hong for her feedbacks on our work. From Hwajung's Data Visualization course, we gained knowledge about the process of visualization design and validation that was utilized in this project.
}

\bibliographystyle{abbrv-doi}

\bibliography{main}
\end{document}